\documentclass[%
 preprint,%
 amssymb, amsmath,nofootinbib%
]{revtex4-1}
\usepackage{epsfig,graphics,graphicx}
\usepackage{color}
\usepackage{bm}%

\newif\iffigure
\figuretrue

\newcommand{\x}{{\bm x}}
\newcommand{\y}{{\bm y}}
\newcommand{\z}{{\bm z}}
\newcommand{\bk}{{\bm k}}
\newcommand{\bb}{{\bm b}}
\newcommand{\br}{{\bm r}}

\def\be{\begin{eqnarray}}
\def\ee{\end{eqnarray}}

\begin{document}

\begin{flushright}
Nikhef-2016-041\\
YITP-16-104
\end{flushright}

\title{Wigner, Husimi and GTMD distributions \\ in the Color Glass Condensate}%

\author{Yoshikazu Hagiwara$^a$, Yoshitaka Hatta$^b$ and Takahiro Ueda$^c$}%
\vspace{3mm}
\affiliation{$^a$Department of Physics, Kyoto University, Kyoto 606-8502, Japan, \\ $^b$Yukawa Institute for Theoretical Physics, Kyoto University, Kyoto 606-8502, Japan, \\ $^c$Nikhef Theory Group, Science Park 105, 1098 XG Amsterdam, The Netherlands }%

\begin{abstract}
We study the phase space distributions of gluons inside a nucleon/nucleus in the small-$x$ regime including the gluon saturation effect. This can be done by using the relation between the gluon Wigner distribution and the dipole S-matrix at small-$x$, the latter satisfies the Balitsky-Kovchegov (BK) equation. By efficiently solving the BK equation with impact parameter dependence, we compute the Wigner, Husimi and generalized TMD (GTMD)  distributions in the saturation regime. We also investigate 
 the elliptic  angular dependence of these distributions which has been recently shown to be measurable in DIS experiments.  

\end{abstract}
\date{\today}

\maketitle


\section{Introduction}

In this paper we explore the phase space distribution of partons inside a high energy nucleon/nucleus. By `phase space' we mean the  five-dimensional space spanned by the longitudinal momentum fraction $x$, the transverse momentum ${\bm k}$, and the impact parameter ${\bm b}$. The corresponding distribution function, the Wigner distribution $W(x,{\bm k},{\bm b})$ \cite{Ji:2003ak,Belitsky:2003nz,Lorce:2011kd,Lorce:2011ni,Mukherjee:2014nya,Mukherjee:2015aja,Liu:2015eqa,Chakrabarti:2016yuw},\footnote{To be precise, the original proposal in \cite{Ji:2003ak,Belitsky:2003nz} was to study the six-dimensional distribution, adding an extra spatial dimension to take account of the skewness dependence in GPDs. The `reduced' five-dimensional form is due to \cite{Lorce:2011kd}.} carries complete information about the single-parton properties of the nucleon, and is often referred to as `Mother distribution' since it 
reduces to  the transverse momentum dependent distribution (TMD) and the Fourier transform of the generalized parton distribution (GPD)  upon integration  over ${\bm b}$ and ${\bm k}$, respectively.

In addition to the Wigner distribution, two associated phase space distributions have been proposed. One is the generalized TMD (GTMD) $F(x,{\bm k},{\bm \Delta})$ \cite{Meissner:2009ww,Lorce:2013pza} (${\bm \Delta}$ is the momentum transfer) which is the Fourier transform of the Wigner distribution with respect to ${\bm b}$. Being fully expressed by momentum variables, the GTMDs are more directly connected to the GPDs, and thus to phenomenology. This makes their classification  \cite{Meissner:2009ww,Lorce:2013pza} and quantum evolution easier to analyze  \cite{Echevarria:2016mrc}. The other is the Husimi distribution $H(x,{\bm k},{\bm b})$ obtained from the Wigner distribution via Gaussian smearing in both ${\bm k}$ and ${\bm b}$ \cite{Hagiwara:2014iya}. Unlike the Wigner distribution,  the Husimi distribution is positive and can be interpreted as a probability distribution in phase space. Moreover, we shall see that the Wigner and GTMD distributions are subject to uncertainties associated with nonperturbative (confinement) effects, whereas the Husimi distribution does not have this  problem.

So far, the studies of these distributions  have been mostly confined to formal theoretical issues and simple model calculations, with little reference to phenomenology. The only exception is a particular GTMD called $F_{14}$ \cite{Meissner:2009ww} which is related to the canonical orbital angular momentum of quarks and gluons in the nucleon \cite{Lorce:2011kd,Hatta:2011ku}, and one therefore has a strong motivation to study it in high energy processes  \cite{Courtoy:2013oaa,Kanazawa:2014nha,Rajan:2016tlg}. In general, however, experimentally measuring the phase space distribution of a quantum system is  a very difficult task. While some successful examples are known in the field of quantum optics (see, e.g., \cite{optics}),  systematic methods in QCD are unfortunately not available. 

This situation recently took an interesting turn when the authors of \cite{Hatta:2016dxp} showed that the {\it gluon} Wigner distribution  for small values of $x$ is experimentally accessible in  diffractive dijet production in DIS (see also \cite{Altinoluk:2015dpi}). This is based on the observation that, at small-$x$ where the gluon saturation becomes important, the Wigner distribution is approximately related to the so-called dipole S-matrix---the forward amplitude of a $q\bar{q}$ pair scattering off a high energy target. It has been further argued that the exclusive measurement of the dijet momenta can reveal  the characteristic angular correlation between ${\bm k}$ and ${\bm b}$.  At small-$x$, this correlation can be written in the form 
\be
W(x,{\bm b},{\bm k}) = W_0 (x,b,k) + 2\cos2(\phi_k-\phi_b) W_1(x,b,k)+\cdots\,.
\ee
The angular dependent term $W_1$ is dubbed `the elliptic Wigner distribution' in \cite{Hatta:2016dxp}. 

 Motivated by these developments, in this paper we compute the gluon Wigner, Husimi and GTMD distributions at small-$x$ including the gluon saturation effect. This is achieved by numerically solving the Balitsky-Kovchegov (BK) equation \cite{Balitsky:1995ub,Kovchegov:1999yj} for the dipole S-matrix keeping the dependence on impact parameter ${\bm b}$. A notable feature of our computation, as compared to previous works on the BK equation with impact parameter \cite{GolecBiernat:2003ym,Ikeda:2004zp,Marquet:2005zf,Berger:2010sh}, is that we assume the hidden SO(3) symmetry of the BK equation postulated by Gubser \cite{Gubser:2011qva}. 
This greatly simplifies the numerics. 
Using this solution, we compute the angular independent and dependent parts of the Wigner distribution, $W_0$ and $W_1$, separately. We then perform additional Gaussian smearings and  Fourier transformations to obtain the Husimi and GTMD distributions. 


\section{Impact parameter dependent dipole S-matrix }\label{Sec.Solve_BK_eq}

\subsection{Hidden symmetry of the BK equation}

Let us first recall the approximate formula of the gluon Wigner distribution at small-$x$ 
derived in \cite{Hatta:2016dxp} 
\be \label{WignerDist}
xW(x,\bk,\bb) = -\frac{2 N_c}{\alpha_S} \int \frac{d^2\br}{(2\pi)^2} e^{i \bk\cdot \br} \left(\frac{1}{4}\nabla^2_\bb +\bk^2 \right) T_Y(\br,\bb)\,,   \label{la}
\ee
where $Y\equiv \ln 1/x$ is the rapidity. The dipole amplitude $T_Y(\br,\bb)=1-S_Y(\br,\bb)$ is related to the dipole S-matrix
\be
S_Y(\br,\bb)= \frac{1}{N_c} \left\langle {\rm Tr}\,  U\left(\bb+\frac{\br}{2}\right)U^\dagger\left(\bb-\frac{\br}{2}\right)\right\rangle_Y\,,
\ee
which is the product of two Wilson lines $U$ representing the forward S-matrix of a quark at $\x=\bb+\br/2$ and an antiquark at $\y=\bb-\br/2$ in the eikonal approximation.  The target averaging $\langle...\rangle_Y$ is done according to the Color Glass Condensate formalism \cite{Gelis:2010nm}.
To leading logarithmic accuracy and in the large $N_c$ limit, the rapidity evolution of $S_Y$ is governed by  the BK equation 
\be
\label{BK_eq}
\partial_Y S_Y(\x,\y) = \frac{\bar{\alpha}_s }{2\pi}  \int d^2\z \frac{(\x-\y)^2}{(\x-\z)^2 (\z-\y)^2} \{S_Y(\x,\z) S_Y(\z, \y) - S_Y(\x,\y)  \}\,,\label{bk}
\ee
where $\bar{\alpha}_s\equiv \frac{N_c \alpha_s}{\pi}$. We shall assume fixed coupling and set $\bar{\alpha}_s =0.2$ throughout this paper.  (We use $(\bb,\br)$ and $(\x,\y)$ interchangeably for the arguments of $S_Y$.)

For our purpose, it is essential to solve (\ref{bk}) keeping the dependence on $\bb = (\x+\y)/2$.  This is numerically expensive, as it involves discretization in $b,r$ and the relative angle $\phi_{b}-\phi_r\equiv \phi_{br}$, but it has been done in the literature with varying degrees of sophistication \cite{GolecBiernat:2003ym,Ikeda:2004zp,Marquet:2005zf,Berger:2010sh}. 
In order to simplify this part of the calculation, following Gubser \cite{Gubser:2011qva}, we  assume that the solution is invariant under certain SO(3) subgroup of the conformal (M$\ddot{\text{o}}$bius) group, the latter being the maximal symmetry of the BK equation in the transverse plane. (See, also, a similar idea in \cite{Hatta:2009nd}.)  
Under this assumption, the solution depends on $\x$ and $\y$ only through the `chordal distance' 
\be
d^2(\x,\y) \equiv \frac{R^2(\x-\y)^2}{(R^2+\x^2)(R^2+\y^2)} = \frac{R^2r^2}{\left(R^2+b^2+\frac{r^2}{4}\right)^2-\frac{b^2r^2}{2} -\frac{b^2r^2}{2}\cos 2\phi_{br}}\,, \label{cho}
\ee
that is, $S_Y(\x,\y)= S_Y(d^2(\x,\y))$.  $R$ is an arbitrary parameter with the dimension of length.  In Appendix \ref{App_proofd2}, we show that $d^2$ satisfies the condition $0\le d^2\le 1$. 

Obviously, this greatly simplifies the numerical calculation, but we have to first argue whether such an assumption makes sense. As a matter of fact,   
  conformal symmetry is rarely exploited in the context of the BK equation (see, however, \cite{Bondarenko:2015fca}) because it is broken by realistic initial conditions. Nevertheless, we conjecture that the SO(3) symmetry, even if it is broken initially, is dynamically restored by the equation. This is based on a curious symmetry found in the numerical results of  \cite{GolecBiernat:2003ym}. There the authors noticed that, after a few units of rapidity evolution, the small-$r$ and large-$r$ regions of $S(\br,\bb)$ become symmetric (cf. Fig.~\ref{fig1} below) even though the initial condition is very asymmetric.  They then commented:  
 ``It is interesting is that the amplitude ($T_Y(\br,\bb)$)  has a maximum for the dipole size which is twice its impact parameter $r=2b$.'' The symmetry between the limits $r\to 0$ and $r\to \infty$ is indicative of conformal symmetry \cite{Hatta:2007fg}. As for the location of the maximum $r=2b$, notice that, for fixed values of $b$ and $\phi_{br}$, (\ref{cho}) is exactly invariant under 
\be
r\to \frac{r_m^2}{r}\,, \qquad r_m=2\sqrt{b^2+R^2}\,,  \label{turn}
\ee
and $r_m \approx 2b$ when $b\gg R$. 
Therefore, we interpret the findings in \cite{GolecBiernat:2003ym}   as a numerical evidence of dynamical SO(3) symmetry restoration.  
Of course, in reality the confinement effect enters when $r\gtrsim r_m$, so the large-$r$ part of the solution is not physically meaningful. Still, the small-$r$ branch of the solution correctly captures the essentials of saturation physics. 
We thus take the following strategy: Since the symmetry is eventually restored, we assume it from the beginning. Specifically, we solve (\ref{bk}) with the initial condition  
\be
S_{Y=0}(\br,\bb) = e^{-d^2(\br,\bb)}\,.
\ee 
(Note that $S_{Y=0}\approx e^{- r^2/R^2}$ when $r,b\ll R$.)  
 We then include confining effects later by hand, when computing the Wigner distribution (\ref{WignerDist}) via Fourier transformation in $\br$.

\subsection{Solving the BK equation with SO(3) symmetry} 

Here we outline how we actually solve (\ref{bk}). An alternative approach is presented in Appendix B. Let us set $\y=0$ after which the equation becomes 
\be
\partial_Y S(\x,0)= \bar{\alpha}_s\int \frac{d^2\z}{2\pi}\frac{x^2}
{(\x-\z)^2 z^2}\left(S(\x,\z)S(\z,0)-S(\x,0)\right)\,. \label{g}
\ee Since $d^2=x^2/(R^2+x^2)$, we can write
\be
S_Y(\x,0) = S_Y\left(\frac{x^2}{R^2+x^2}\right) \equiv g_Y(x)=g_Y\left(\sqrt{\frac{R^2d^2(\x,0)}{1-d^2(\x,0)}}\right)\,.
\ee
If we know the function $g_Y(x)$, we immediately get 
\be
S_Y(\x,\y)=S_Y(d^2(\x,\y))=g_Y\left(\sqrt{\frac{R^2d^2(\x,\y)}{1-d^2(\x,\y)}}\right)\,. 
\label{solution}
\ee
Noting that
\be
S_Y(\x,\z)=S_Y\left(\frac{R^2(\x-\z)^2}{(R^2+x^2)(R^2+z^2)}\right) =g_Y\left(\sqrt{\frac{R^4(\x-\z)^2}{R^4+x^2z^2+2R^2\x\cdot \z}}\right)\,,
\ee
we can recast (\ref{g}) into  an equation for $g_Y(x)$ 
\be
\partial_Y g_Y(x) = \bar{\alpha}_s  \int \frac{d^2\z}{2\pi}\frac{x^2}
{(\x-\z)^2 z^2}\left\{g_Y\left(\sqrt{\frac{R^4(\x-\z)^2}{R^4+x^2z^2+2R^2\x \cdot \z}}\right)g_Y(z)-g_Y(x)\right\} \,. \label{gg}
\ee
Since the left hand side is independent of the angle of $\x$, we can set $\phi_x=0$ and arrive at
\be
\partial_Y g(x) &=&  \bar{\alpha}_s\int_0^{2\pi} \frac{d\phi}{2\pi} \int_0^\infty \frac{dz}{z} 
\frac{x^2} 
{(x^2+z^2-2xz\cos\phi)} \nonumber \\ 
&& \qquad\times \left\{g_Y\left(\sqrt{\frac{R^4(x^2+z^2-2xz\cos\phi)}{R^4+x^2z^2+2R^2xz \cos\phi}}\right)g_Y(z)-g_Y(x)\right\}\,. \label{to}
\ee
We solved (\ref{to}) numerically  with $R=1$ and the initial condition
\be
g_{Y=0}(x)=e^{- d^2(\x,0)} = \exp\left(-\frac{x^2}{R^2+x^2}\right)\,,
\ee
and obtained $S_Y(d^2)$ from (\ref{solution}). 
The result is shown in the left panel of Fig.~\ref{fig1} at different values of $Y$ up to $Y=10$. In the right panel, we show $T_Y=1-S_Y$ as a function of $\ln r$ at fixed $b=1$ and $\cos  \phi_{br}=0$. As expected, the peak position is always at $r=r_m = 2\sqrt{b^2+R^2}=2\sqrt{2}$ (see (\ref{turn})), irrespective of the value of $Y$. On the other hand,  the {\it saturation momentum} $Q_s(Y,b,\phi_{br})$, defined by the condition $T_Y(r=1/Q_s<r_m)=const.$, grows  with $Y$. The $r>r_m$ part of the solution is unphysical  and should not affect physical observables.  

\iffigure
\begin{figure}[t]
 \includegraphics[width=83mm]{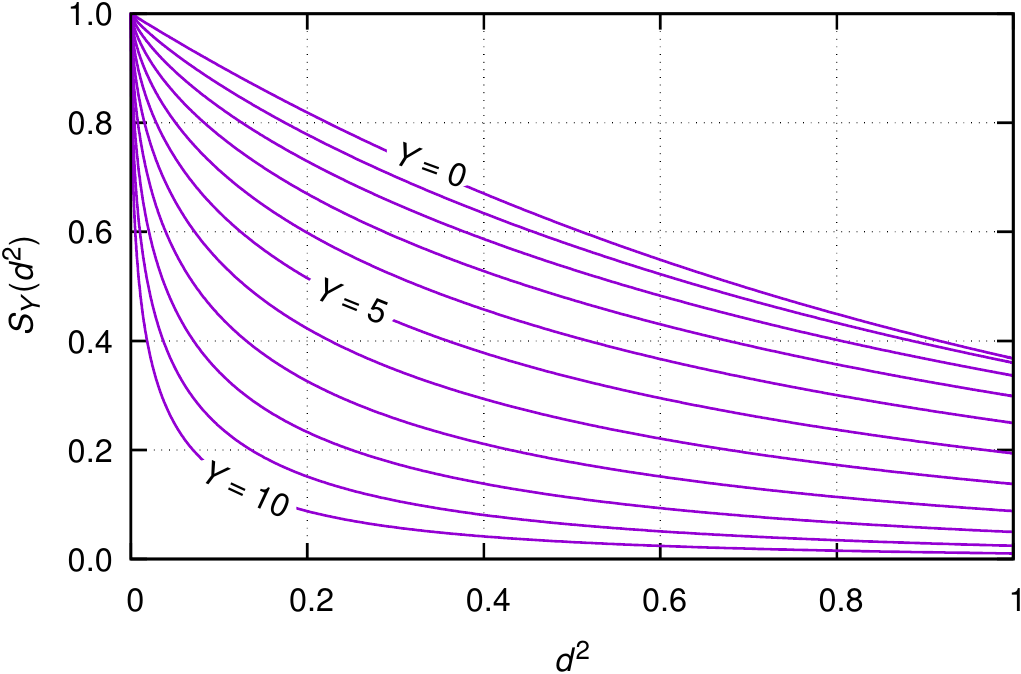}
 \includegraphics[width=80mm]{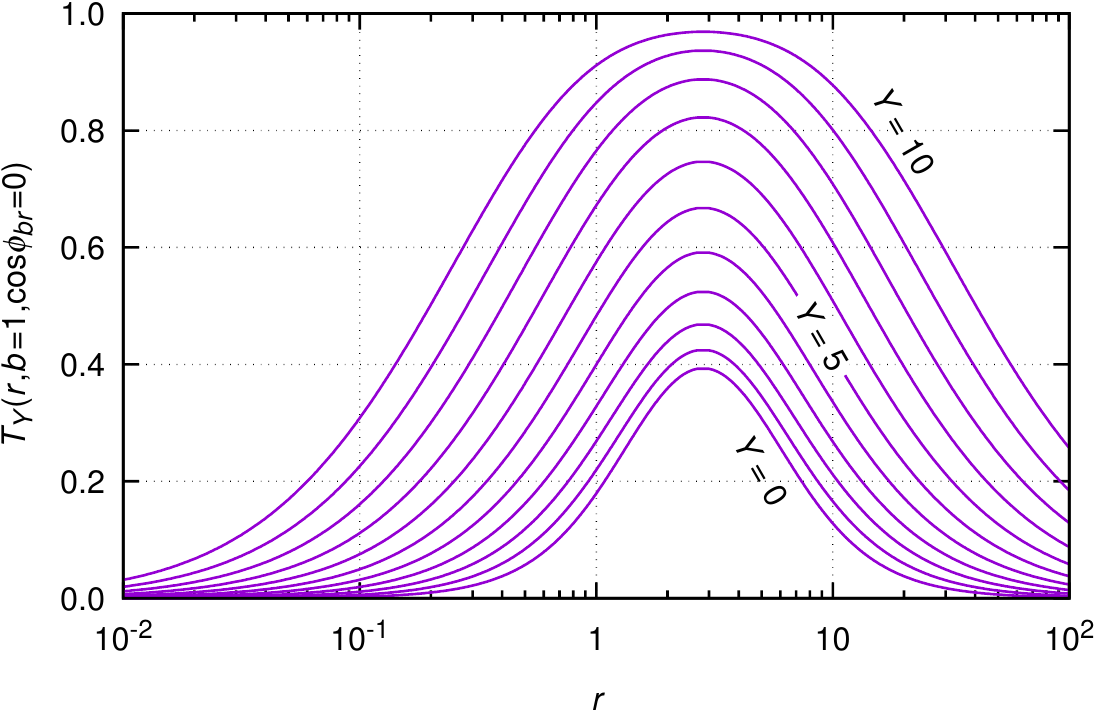}
\caption{Left:  Dipole S-matrix $S_Y(d^2)$ as a function of $d^2$ up to $Y=10$. Right: $T_Y=1-S_Y$ as a function of $\ln r$ at  $b=1$ and $\cos(\phi_b- \phi_r)=0$. } \label{fig1}
\end{figure}
\fi


\section{Wigner Distribution}\label{Sec.Wigner}

Now that we have a solution $T_Y(\br,\bb)$ of the BK equation, it should be straightforward to perform the Fourier transform in (\ref{la}) to obtain the Wigner distribution. However, this does not produce a physical result due to the following reason.  It is known that the small-$r$ behavior of $T_Y$ takes the `geometric scaling' form
\be
T_Y(\br,\bb) \propto (rQ_s)^{2\gamma}\,. \quad (r\to 0) \label{cf}
\ee
 The exponent is $\gamma=1$ initially, but with increasing $Y$ it becomes weakly $r$-dependent and interpolates between  the `saturation anomalous dimension'  $\gamma\approx 0.63$ around $r \lesssim 1/Q_s$ \cite{Gribov:1984tu,Iancu:2002tr} and the asymptotic value $\gamma=1$ as $r\to 0$. (We found $\gamma \approx 0.79$ at $Y=5$ and $\gamma\approx 0.73$ at $Y=10$ around $r\sim 10^{-4}$.)   Due to the SO(3) symmetry, the small-$r$ and large-$r$ behaviors are related so that 
$T_Y(r) \sim 1/r^{2\gamma}$ in the large-$r$ region. 
The $r$-integral in (\ref{la}) then becomes, after integrating over the azimuthal angle,
\be
\int^\infty dr \frac{rJ_0(kr)}{r^{2\gamma}}\,.
\ee
This is a convergent integral (for $\gamma>0.25$), but it converges slowly due to the oscillation of the Bessel function. It is pointless to try to perform this integral accurately  because the perturbative tail at large distances ($r\gg r_m$) is unphysical and should not affect physical observables.\footnote{Nevertheless, we performed the integral (\ref{la}) as it is. The result is that there are two peaks in the $k$-direction, one at $k\sim Q_s$, which is physical, and the other at $k\sim 1/(r_m^2Q_s)$ which is totally an artifact of conformal symmetry.} Note that this problem is not an artifact of our assumption of SO(3) symmetry. The same problem should appear for the solution in \cite{GolecBiernat:2003ym}, and including higher-order corrections, such as the running coupling effect \cite{Berger:2010sh},  will not help solve the problem. Rather, it is an artifact of the BK equation itself whose kernel features the perturbative Coulomb interaction at large distances.   What is expected to occur in real QCD is that $T_Y$ approaches the black disc limit $T_Y(r\gg R)\to 1$ due to confinement, and the large-$r$ region of the integral (\ref{la}) gives a vanishing contribution $\delta^{(2)}(\bk)\bk^2=0$. However, it is difficult to properly implement the effect of confinement directly in the BK equation (see an attempt in \cite{Ikeda:2004zp}). 
 
 An elegant way to avoid this problem is to calculate  instead the Husimi distribution in which the $r$-integral is effectively cut off by the built-in Gaussian factor. This will be done in the next section. As for the Wigner distribution, here we show the result obtained in an {\it ad hoc} way, by inserting a Gaussian damping factor by hand. Namely, we compute, instead of (\ref{la}), 
\be \label{Wigner2}
xW'(x,\bk,\bb) = -\frac{2 N_c}{\alpha_S} \int \frac{d^2\br}{(2\pi)^2} e^{i \bk\cdot \br }e^{-\epsilon r^2} \left(\frac{1}{4}\nabla^2_\bb +\bk^2 \right) T_Y(\br,\bb)\,. \label{com}
\ee
 We choose $\epsilon = 1/4$ so that the  region $r\gtrsim 2$ is suppressed. (Remember that $r_m \approx 2R=2$ for small $b$.) 

Let us evaluate the angular independent and dependent parts of $W'$ separately. As is clear from (\ref{cho}), the Fourier expansion of $W'$ contains only even harmonics 
\be
xW'(x,\bk,\bb) = xW_0(x,k,b) + 2 \sum_{n=1}^{\infty} xW_n(x,k,b) \cos(2n \phi_{bk}) \,.
\ee
We only consider the leading term $W_0$, and the `elliptic' term $W_{n=1}$ which is expected to give the dominant angular dependence \cite{Hatta:2016dxp}. They can be isolated as 
\be
xW_0(x,k,b) &=& -\frac{N_c}{2\alpha_S \pi^2}\left(\frac{1}{4} \frac{\partial^2}{\partial b^2} + \frac{1}{4b}\frac{\partial}{\partial b}  +k^2 \right)  \int_0^{\infty} r  e^{-\epsilon r^2}J_0(kr)dr  \nonumber \\ 
&& \qquad \times \int_0^{2\pi} d\phi_{br} T_Y(r,b,\cos2\phi_{br}) \,,
\ee
and
\be
xW_1(x,k,b) &=& \frac{N_c}{2\alpha_S \pi^2} \left(\frac{1}{4} \frac{\partial^2}{\partial b^2} + \frac{1}{4b}\frac{\partial}{\partial b}  -\frac{1}{b^2}+k^2 \right) \int_0^{\infty} r e^{-\epsilon r^2} J_2(kr)  dr  \nonumber \\ 
&& \times \int_0^{2\pi} d\phi_{br} \cos(2\phi_{br})  T_Y(r,b,\cos2\phi_{br}) \,.
\ee
 The results for $W_0$ and $W_1$  are shown in Fig.~\ref{fig2} and Fig.~\ref{fig3}, respectively. 
\iffigure
\begin{figure}[t]
 \includegraphics[width=80mm]{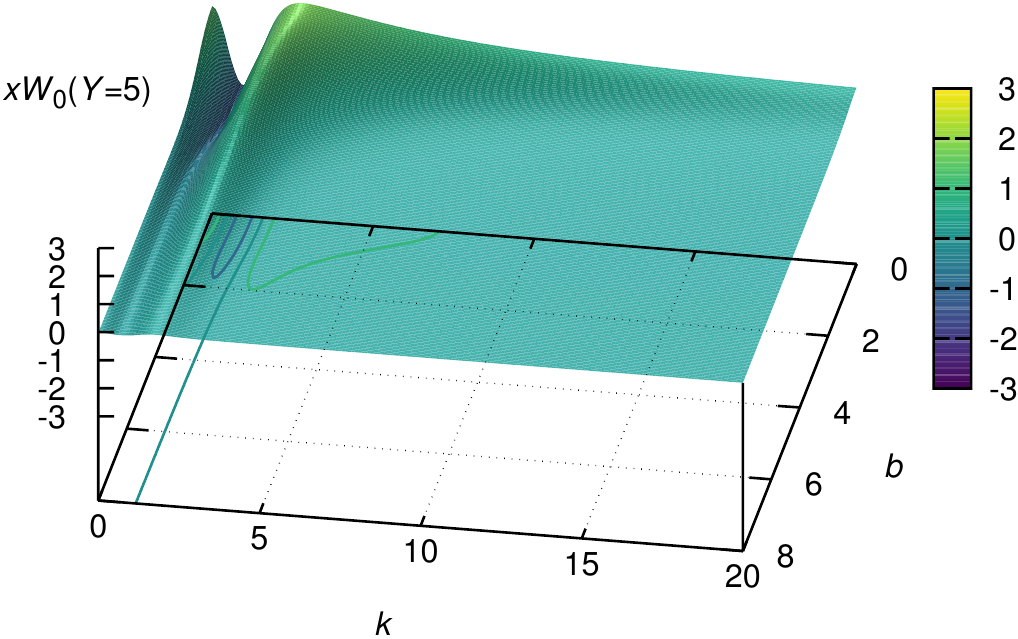}
 \includegraphics[width=80mm]{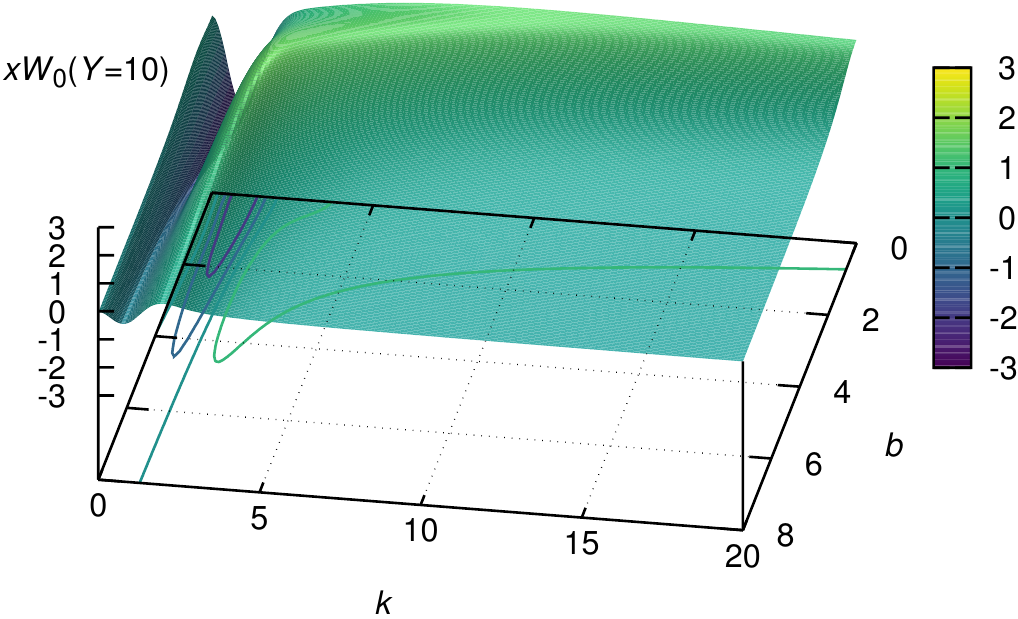}
\caption{ The angular-independent part of the Wigner distribution $xW_0$ in the $(k,b)$ plane. Left: $Y=5$; Right: $Y=10$. } \label{fig2}
\end{figure}
\begin{figure}[htb]
  \includegraphics[width=80mm]{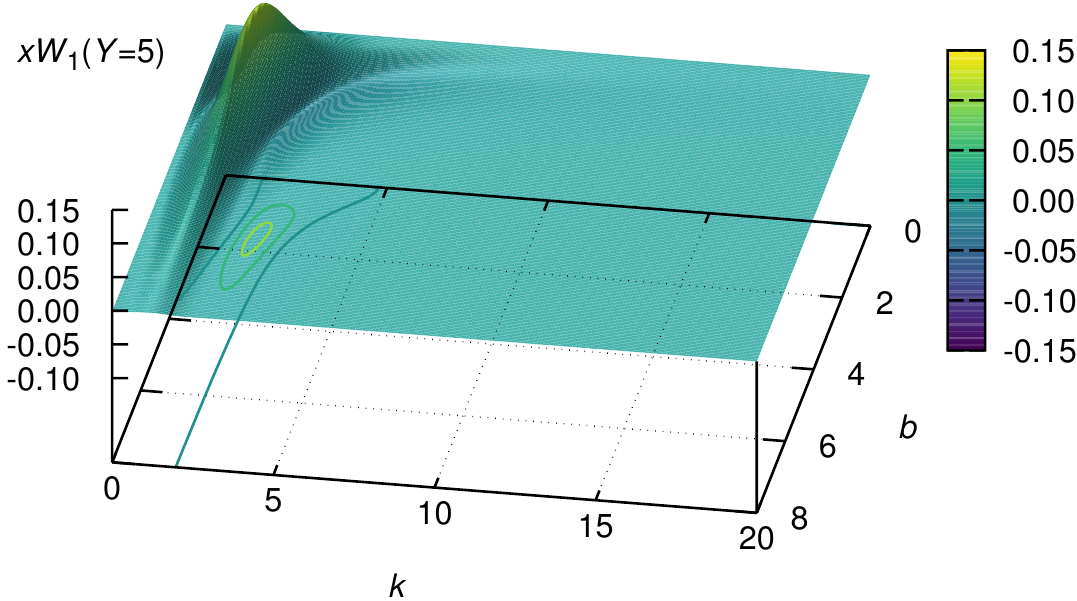}
 \includegraphics[width=80mm]{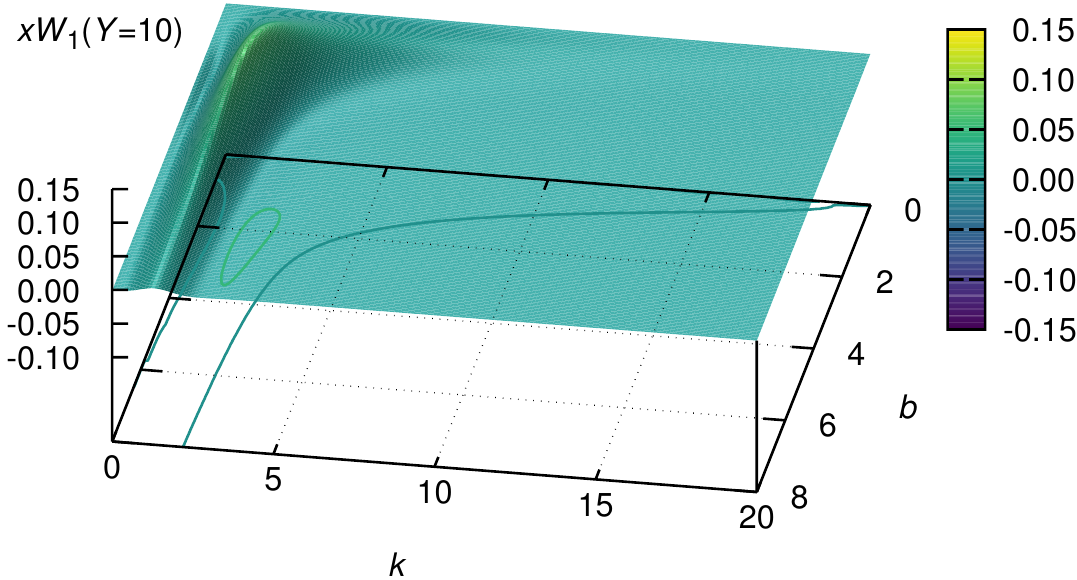}
\caption{ The elliptic Wigner distribution $xW_1$ in the $(k,b)$ plane. Left: $Y=5$; Right: $Y=10$. } \label{fig3}
\end{figure}
\fi
In Fig.~\ref{fig4}, we plot $W_0$ and $W_1$ as a function of $k$ at fixed $b=1$. The peak position of $W_0$ can be identified with the saturation momentum $k=Q_s(Y,b)$ which is an increasing function of $Y$ and a decreasing function of $b$.\footnote{We note that the peak position (normalization of $Q_s$) depends on the Gaussian parameter $\epsilon$.}  The peak of the elliptic part is about $3\sim 5\%$ of that of $W_0$ in magnitude, and interestingly, it moves much more slowly with $Y$. This can be understood as follows.  The SO(3) symmetry implies that in the geometric scaling region $S_Y$ takes the form \cite{Gubser:2011qva}
\be
S_Y(\br,\bb) \sim f(Q^2_s(Y)d^2(\bb,\br))\,. \label{geo}
\ee
In the small-$r$ region such that $r\ll R$ and $b\lesssim R$, one has 
\be
d^2\approx \frac{r^2}{R^2}\left(1+\frac{b^2r^2}{2R^4}\cos 2\phi_{br}\right)\,.
\ee
Because of the extra factor of $r^2$, the angular dependent part cannot show geometric scaling. A naive estimate would be $k_{peak}\sim 1/r \sim \sqrt{Q_s(Y)}$, which is indeed a slower increase with $Y$, but a larger window in $Y$ is needed to really test this behavior. 
 Finally, $W_0$ becomes negative just behind the peak. This is acceptable because the Wigner distribution is not necessarily a positive function. However, it remains to see whether other  more realistic regularization schemes lead to similar conclusions.  

\iffigure
\begin{figure}[h]
 \includegraphics[width=80mm]{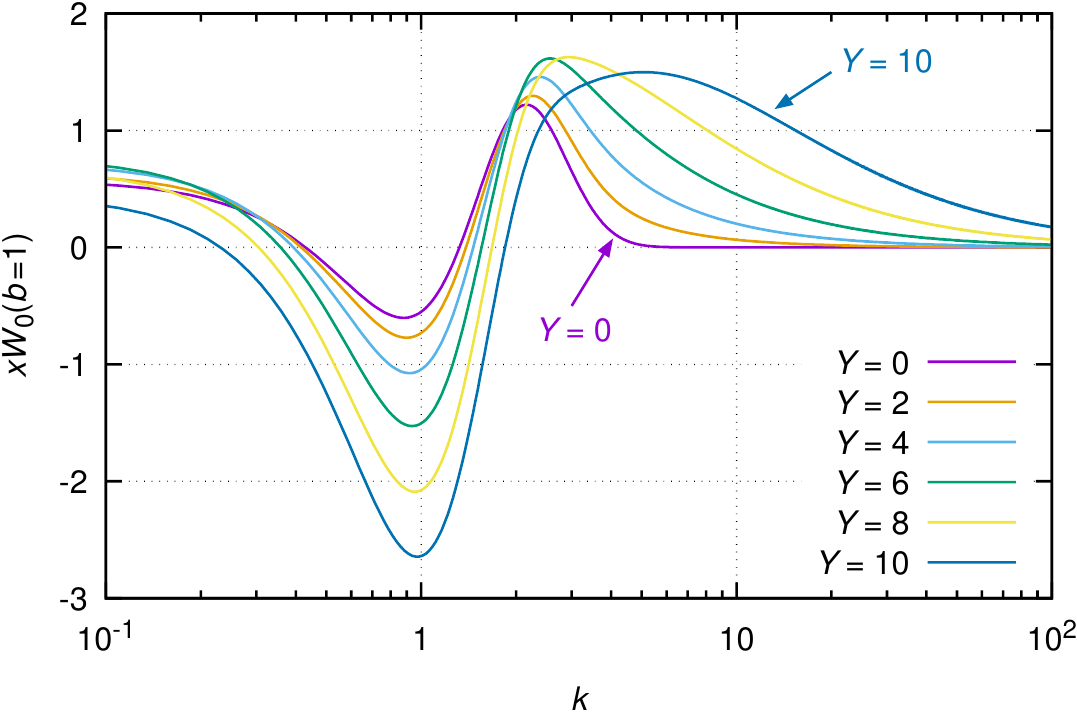}
 \includegraphics[width=80mm]{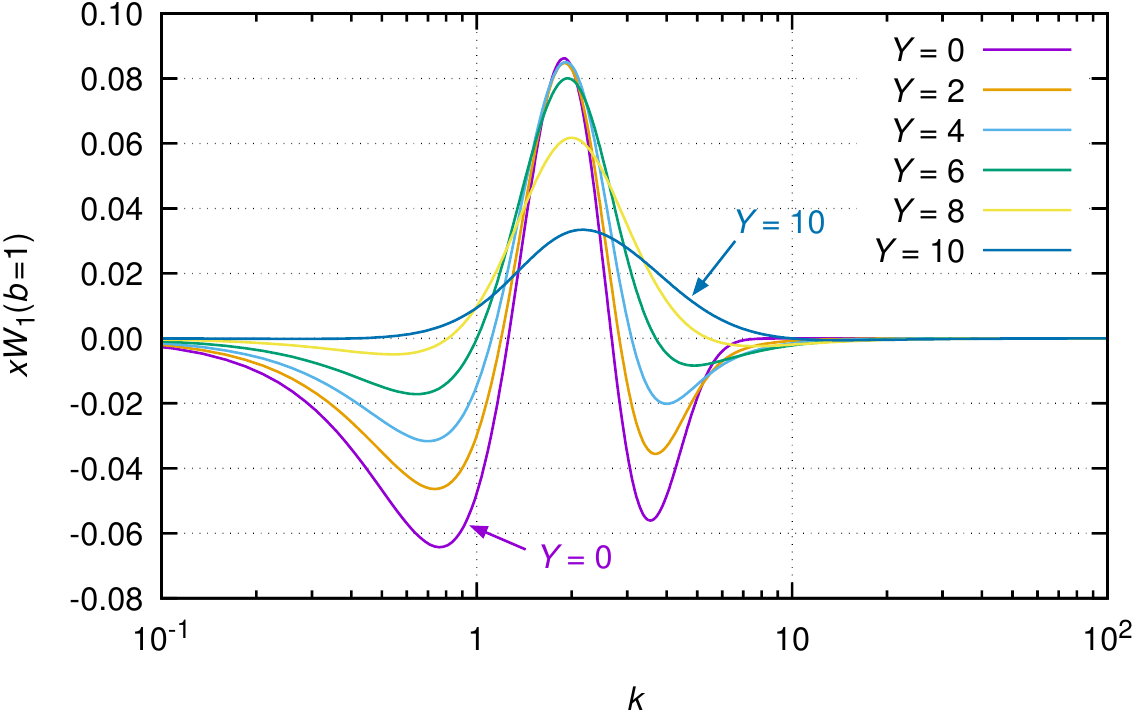}
\caption{ The $k$-distribution of $W_0$ and $W_1$ at fixed $b=1$. } \label{fig4}
\end{figure}
\fi 

\section{Husimi Distribution}\label{Sec.Husimi}

The QCD Husimi distribution is obtained from the Wigner distribution via Gaussian smearing in both $\bk$ and $\bb$  \cite{Hagiwara:2014iya} 
\be
xH(x,\bk,\bb) := \frac{1}{\pi^2}\int d^2\bb' d^2\bk' e^{-\frac{1}{l^2}(\bb-\bb')^2 - l^2(\bk-\bk')^2} xW(x,\bk',\bb') \,. \label{husimi}
\ee
Note that the widths of the two Gaussian factors are inversely related so that they obey the minimum uncertainty relation $\delta k \delta b=\frac{1}{2}$. In nonrelativistic quantum mechanics, this condition ensures that the Husimi distribution is positive semi-definite $H\ge 0$. 
From (\ref{WignerDist}), we obtain  
\be
xH(x,\bk,\bb) &=& -\frac{2N_c}{l^4\alpha_S\pi} \int d^2\bb' \frac{d^2\br}{(2\pi)^2} e^{-\frac{1}{l^2} (\bb - \bb')^2 -\frac{\br^2}{4l^2} +i\bk\cdot \br} \nonumber \\
 && \qquad \times \left\{ \frac{1}{l^2} (\bb - \bb')^2 + l^2\left(\bk + \frac{i\br}{2l^2} \right)^2  \right\} T_Y(\br,\bb') \,, 
\ee
where we integrated by parts in $\bb'$. Thanks to the Gaussian factors, the integrals converge rapidly. Performing integrations over the azimuthal angles, we arrive at
\be
xH_0(x,k,b) &=& -\frac{2N_c}{l^4\alpha_S \pi} \int b' db' \frac{rdr}{2\pi} e^{-\frac{1}{l^2} (b^2 + {b'}^2) -\frac{r^2}{4l^2} } \nonumber \\
&& \ \ \ \ \ \times \biggl[\left \{ \left(\frac{1}{l^2} (b^2 + {b'}^2) + l^2k^2 - \frac{r^2}{4l^2} \right)I_0\left(\frac{2bb'}{l^2} \right)    - \frac{2b b'}{l^2} I_1\left(\frac{2bb'}{l^2} \right) \right\}J_0(kr)  \nonumber \\ 
&& \ \ \ \ \ \ \ \ \  -kr I_0\left(\frac{2bb'}{l^2} \right) J_1(kr)   \biggr]  \ \int_0^{2\pi} d\phi_{b'r} T_Y(r,b',\cos2\phi_{b'r}) \,, \label{EH_wrong}
\ee
\be
xH_1(x,k,b) &=& \frac{2N_c}{l^4\alpha_S \pi} \int b' db' \frac{rdr}{2\pi} e^{-\frac{1}{l^2} (b^2 + {b'}^2) -\frac{r^2}{4l^2} } \nonumber \\
&& \ \ \ \ \ \times \biggl[\left \{ \left(\frac{1}{l^2} (b^2 + {b'}^2) + l^2k^2 - \frac{r^2}{4l^2} \right)I_2\left(\frac{2bb'}{l^2} \right)    - \frac{2b b'}{l^2} I_1\left(\frac{2bb'}{l^2} \right) \right\}J_2(kr)  \nonumber \\ 
&& \ \ \ \ \ \ \ \ \  +kr I_2\left(\frac{2bb'}{l^2} \right) J_1(kr)   \biggr]  \ \int_0^{2\pi} d\phi_{b'r} \cos 2\phi_{b'r}T_Y(r,b',\cos2\phi_{b'r}) \,. \label{EH}
\ee
The parameter $\ell$ is arbitrary, but here we set $\ell=R=1$ so that  the Gaussian factor in the $r$-integral becomes identical to that in  (\ref{com}),  $\frac{1}{4\ell^2}=\epsilon = \frac{1}{4}$. The result is shown in Fig.~\ref{fig5} for $Y=8$ and $Y=10$. Up to $Y\sim 6$, there is a single peak at the origin of the phase space. The would-be peak at $k=Q_s$ is  covered up. The latter starts to show up around $Y \sim 7$, and becomes a distinct peak for $Y\gtrsim 8$. From that on, it moves towards the larger $k$-region as in the Wigner case. The elliptic part is very small and the peak position does not change appreciably with increasing $Y$. 

We see that the Husimi distribution is everywhere positive (up to numerical errors),  so it can be legitimately interpreted as a probability distribution in phase space.  Actually,  the positivity of the QCD Husimi distribution as defined in (\ref{husimi}) has not been proven. However, as explicitly demonstrated here and also in Ref.~\cite{Hagiwara:2014iya},  in practice one does obtain a positive distribution even though the corresponding Wigner distribution is not necessarily positive.  In  quantum mechanics, the positivity of the Husimi distribution is related to the coherent state which provides the classical-like description of a quantum state. The foundation of the Color Glass Condensate is also the coherent state (i.e., classical gauge fields) \cite{Gelis:2010nm}. Thus the use of the Husimi distribution may be more natural in the small-$x$ saturation regime than in the large-$x$ regime.

\iffigure
\begin{figure}[h]
 \includegraphics[width=80mm]{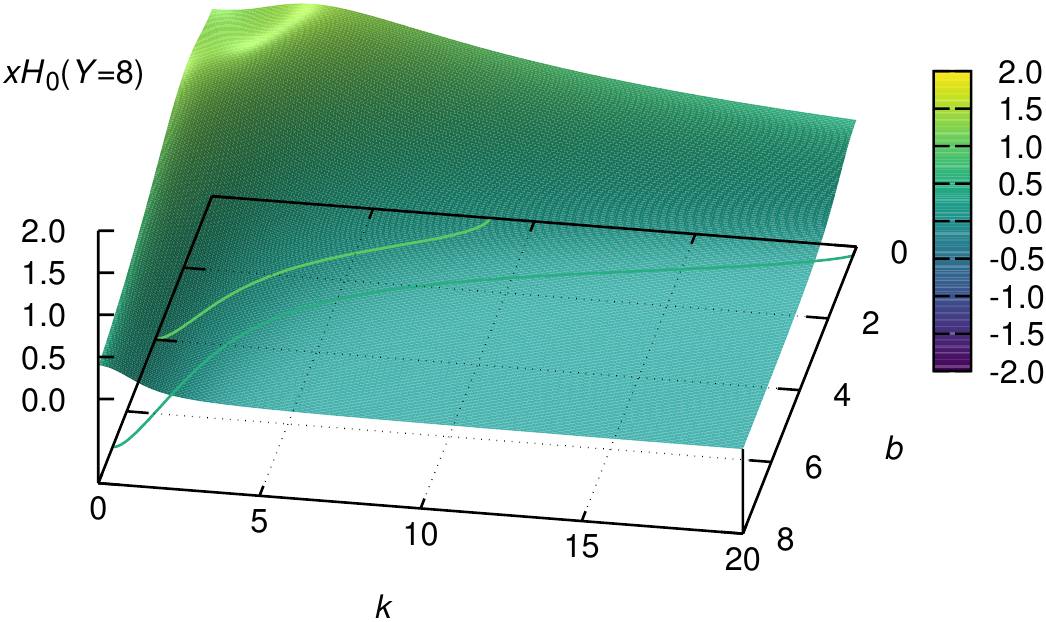}
 \includegraphics[width=80mm]{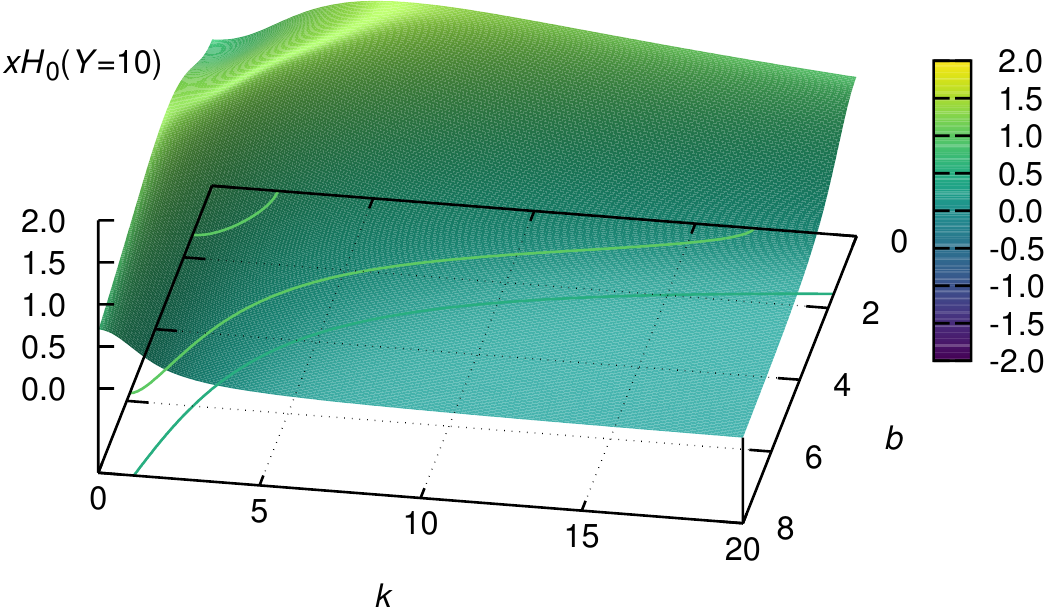}
\caption{ The Husimi distribution at $Y=8$ (left) and $Y=10$ (right). } \label{fig5}.
\end{figure} 
\fi

\iffigure
\begin{figure}[h]
 \includegraphics[width=80mm]{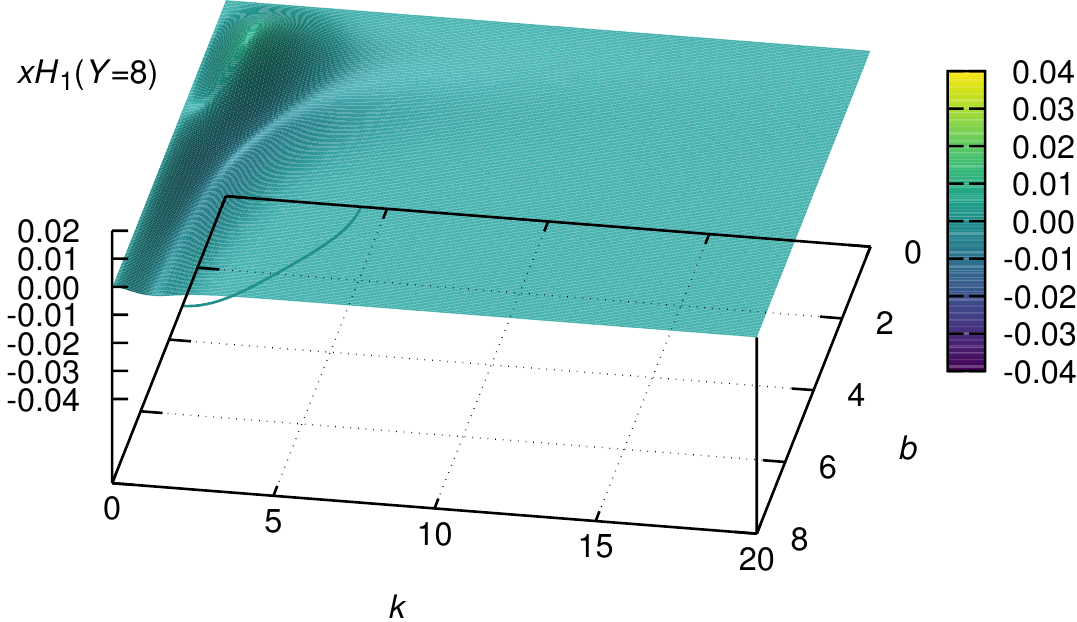}
 \includegraphics[width=80mm]{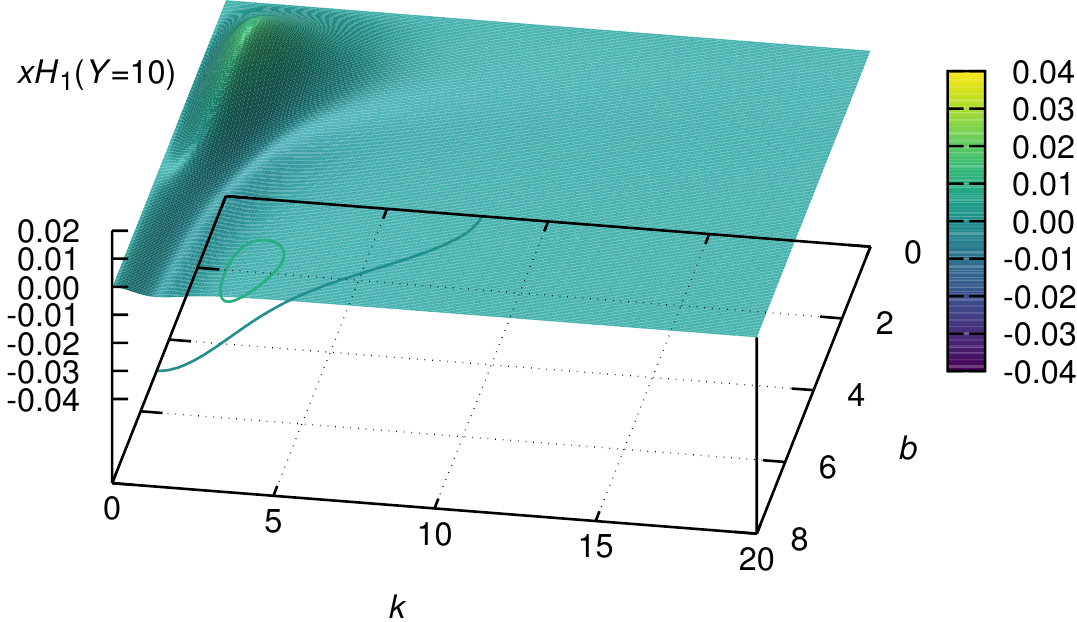}
\caption{ The elliptic Husimi distribution at $Y=8$ (left) and $Y=10$ (right). } \label{fig5}.
\end{figure} 
\fi

\section{Generalized TMD (GTMD) Distribution}\label{Sec.GTMD}
Finally, we consider the gluon  GTMD which is defined as the Fourier transform of the Wigner distribution with respect to $\bb$
\be
x F(x,\bk,{\bm \Delta}) &\equiv& \int \frac{d^2\bb}{(2\pi)^2}\,  e^{i\bb\cdot {\bm \Delta}} x W(x,\bk,\bb) \nonumber \\
&=& \frac{2N_c}{\alpha_s } \left(\frac{\Delta^2}{4} - k^2 \right) \int \frac{d^2 \br}{(2\pi)^2} \int \frac{d^2\bb}{(2\pi)^2} \, e^{i\bk\cdot\br } e^{i\bb \cdot {\bm \Delta}} T_Y(\br,\bb)\,.
\ee
 As before, we can expand $F$ in Fourier harmonics 
\be 
x F(x,\bk,{\bm \Delta})  = xF_0(x,k,\Delta) + 2\cos (2\phi_{k\Delta}) x F_1(x,k,\Delta)+\cdots. 
\ee 
The first two terms can be computed as 
\be
xF'_0(k,\Delta)=   \frac{N_c}{2\pi^2 \alpha_s}\left(\frac{\Delta^2}{4}-k^2\right) \int_0^\infty rJ_0(kr) e^{-\epsilon r^2} dr \int_0^\infty b J_0(b\Delta) db \int \frac{d\phi_{br}}{2\pi} T_Y(\br,\bb)\,, \label{f0}
\ee
\be
xF'_1(k,\Delta) &=&  - \frac{N_c}{2\pi^2 \alpha_s}\left(\frac{\Delta^2}{4}-k^2\right) \int_0^\infty rJ_2(kr) e^{-\epsilon r^2} dr \int_0^\infty b J_2(b\Delta) db  \nonumber \\ 
&& \qquad \times \int \frac{d\phi_{br}}{2\pi} \cos (2\phi_{br}) T_Y(\br,\bb)\,,
\ee
 where again we inserted  a Gaussian factor $e^{-\epsilon r^2}$ due to the same reason as in the Wigner case.  We could have inserted a similar Gaussian factor for the $b$-integral as well, but we decided not to because  the convergence of the $b$-integral is better than that of the $r$-integral. (Note that $T_Y(r,b) \sim (r^2/b^{4})^\gamma$ as $b\to \infty$, cf. (\ref{cf}).)  The results are shown in Fig.~\ref{fig6} and Fig~\ref{fig7}. 
 We see that there is a peak in $xF_0$ when $\Delta$ is small and its height increases rapidly with $Y$. From this, we can define the saturation momentum $k=Q_s(Y,\Delta)$. 
 Note that $xF_0$ falls steeply with $\Delta$ and becomes very small already when $\Delta=1$.  
  In contrast, the elliptic part is peaked at a finite value $\Delta \sim 1$. The $Y$-evolution of the peak is shown in Fig.~\ref{fig8} and Fig.~\ref{fig9}. We see that $Q_s$ is an  increasing function of $\Delta$. This is  consistent with the result in \cite{Marquet:2005zf} and is natural given  that $Q_s(Y,b)$ is a decreasing function of $b$. On the other hand, again the peak position of the elliptic part moves very slowly with $Y$. 

\iffigure
\begin{figure}[h]
 \includegraphics[width=80mm]{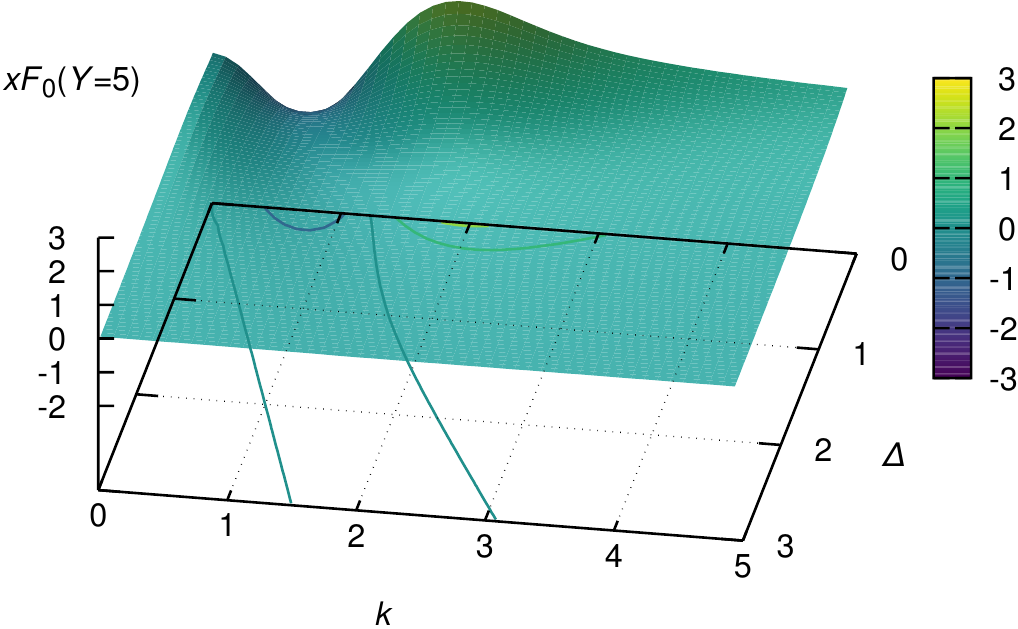}
 \includegraphics[width=80mm]{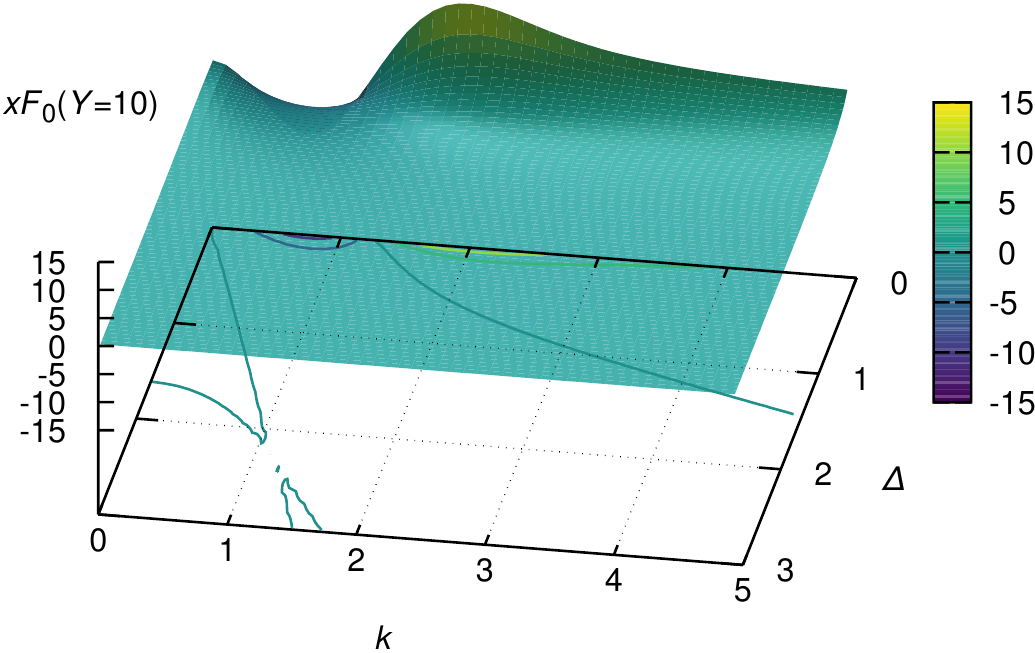}
\caption{ The angular independent part of the GTMD in the $(k,\Delta)$ plane at $Y=5$ (left) and $Y=10$ (right). } \label{fig6}
\end{figure}
\fi 

\iffigure
\begin{figure}[h]
 \includegraphics[width=80mm]{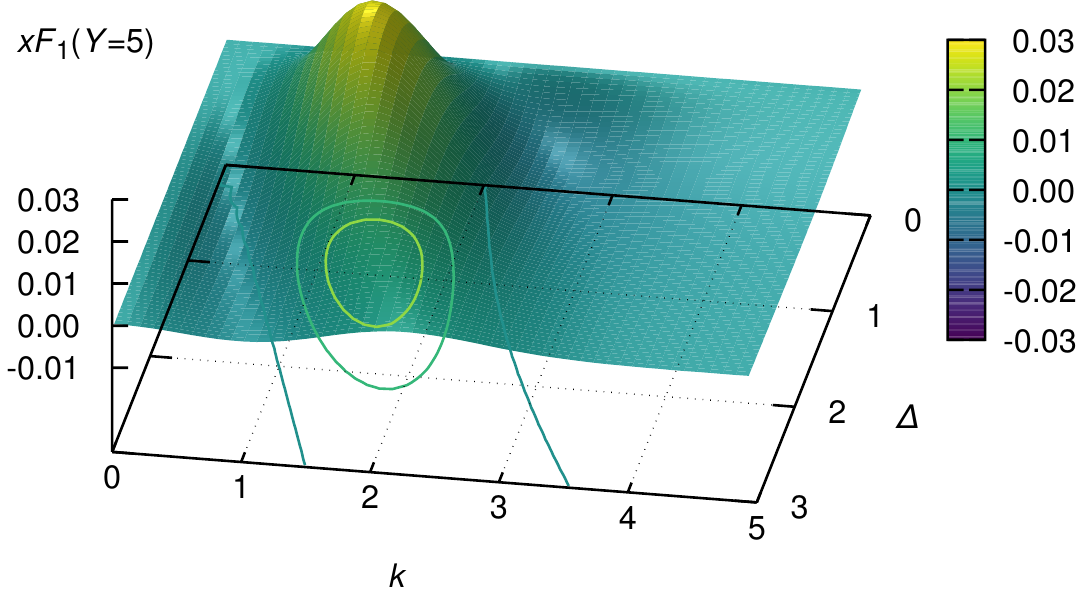}
 \includegraphics[width=80mm]{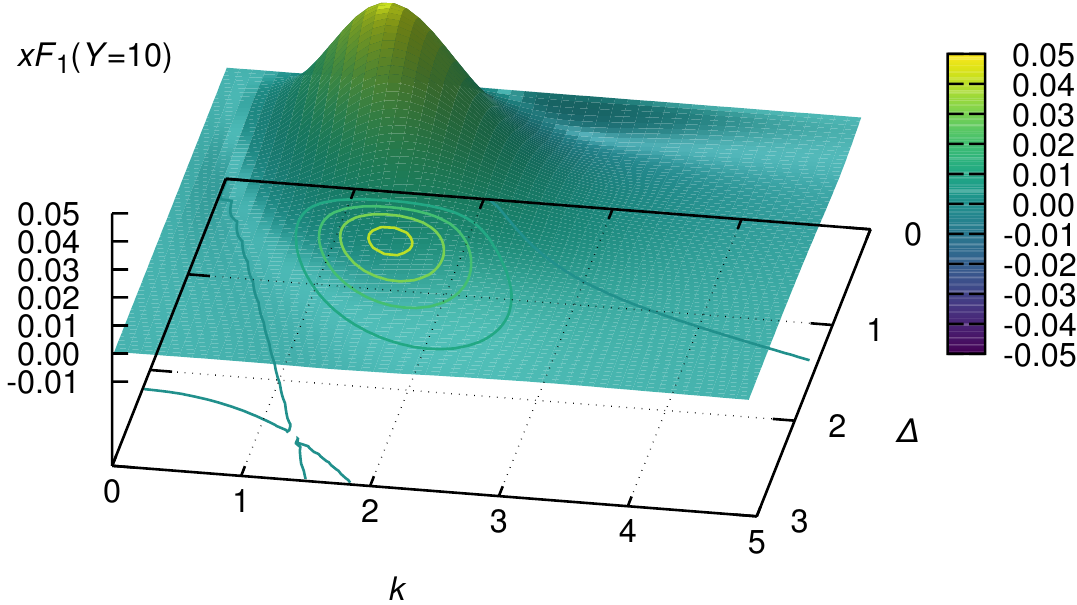}
\caption{The elliptic GTMD in the $(k,\Delta)$ plane at $Y=5$ (left) and $Y=10$ (right). } \label{fig7}
\end{figure}
\fi

\iffigure
\begin{figure}[h]
 \includegraphics[width=80mm]{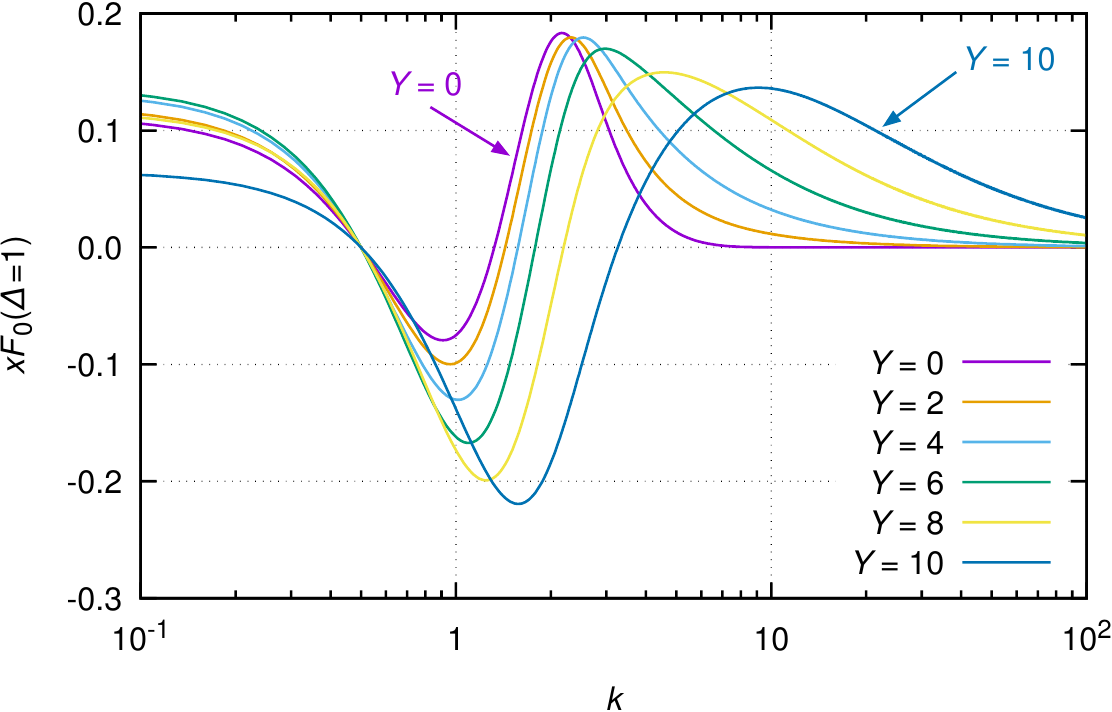}
 \includegraphics[width=80mm]{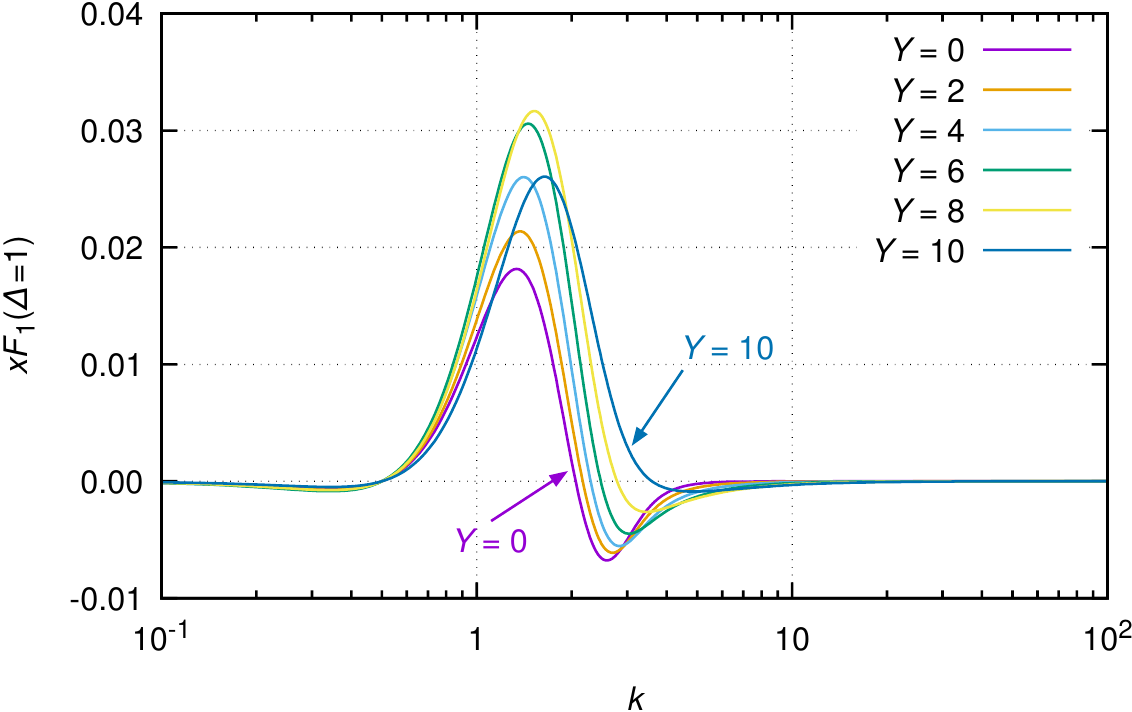}
\caption{ The $Y$-evolution of the peak at $\Delta=1$. Left: angular independent part; Right: Elliptic part. } \label{fig8}
\end{figure}
\fi

\iffigure
\begin{figure}[h]
 \includegraphics[width=80mm]{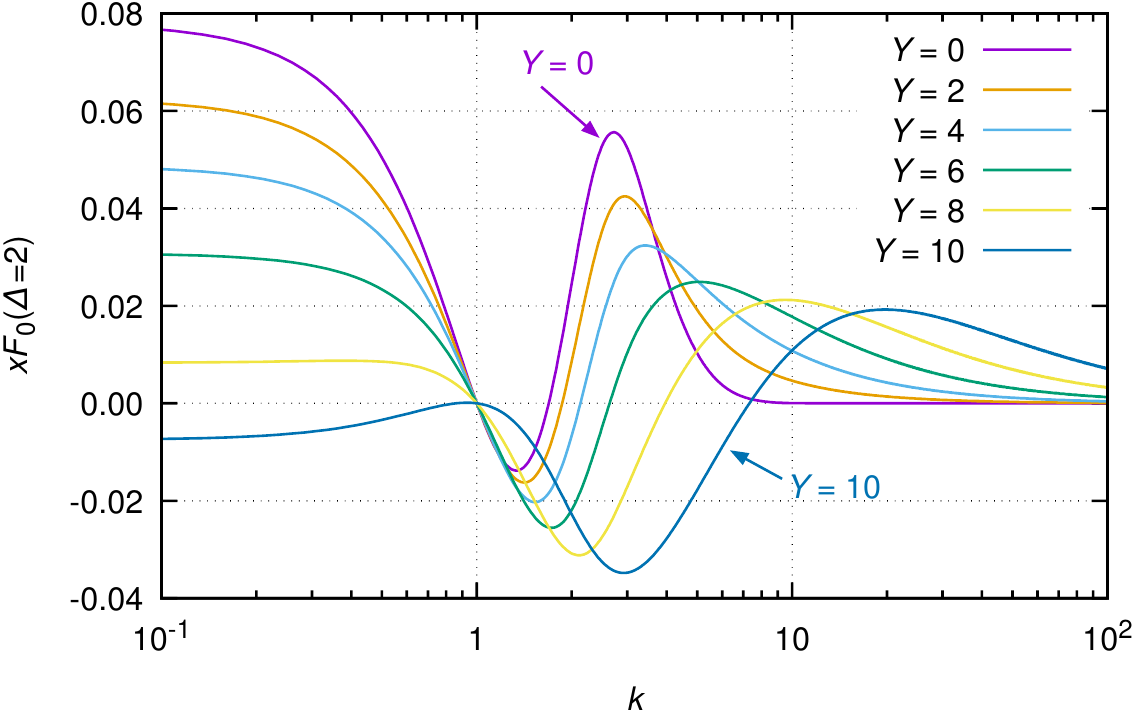}
 \includegraphics[width=80mm]{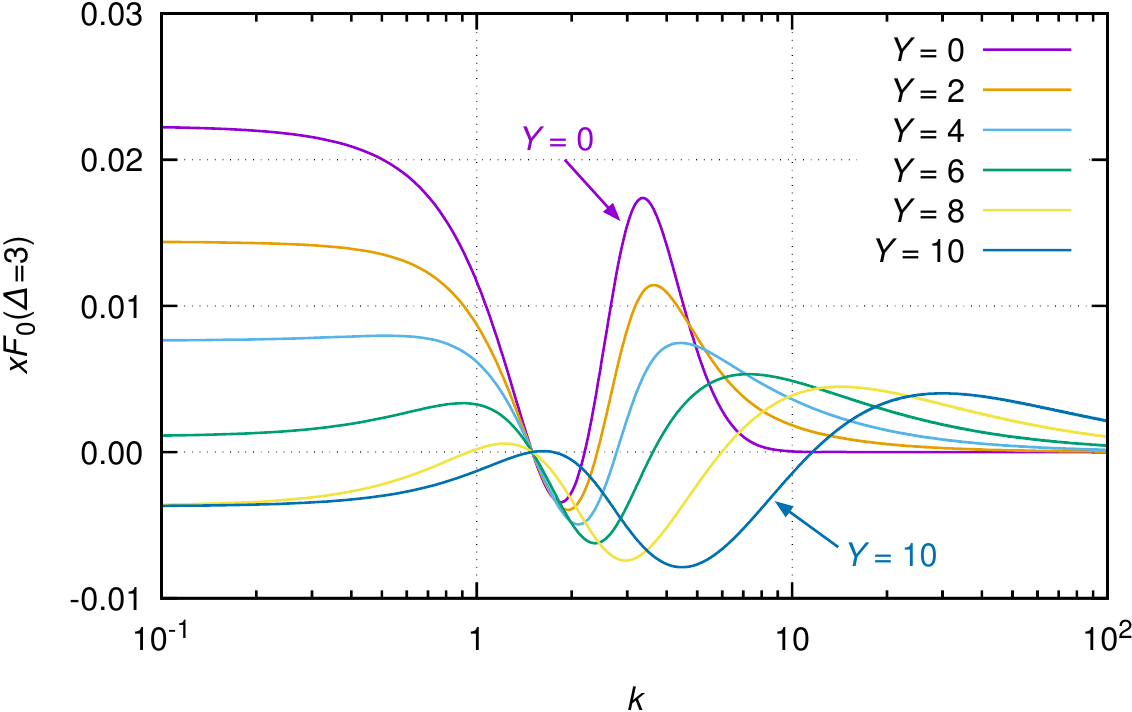}
\caption{ The $Y$-evolution of the peak of $F_0$ at $\Delta=2$ (left) and $\Delta=3$ (right). } \label{fig9}
\end{figure}
\fi 

\section{Conclusion}\label{conclusion}
In this paper we have studied the Wigner, Husimi  and GTMD distributions at small-$x$ including  the gluon saturation effect. To calculate these distributions, we proposed an efficient way to solve the BK equation with impact parameter. This is to exploit the SO(3) symmetry of the equation following \cite{Gubser:2011qva}. We argued that this symmetry is dynamically restored by the equation even if the initial condition is not symmetric. 


We have  seen that the Wigner distribution is sensitive to how we implement  confinement effects in the BK equation, a subject poorly understood. 
 We introduced an {\it ad hoc} Gaussian factor, but then what has been computed is something between the Wigner distribution and the Husimi distributions. For the latter, the Gaussian factors are a part of the definition and come from a well-motivated physical argument. As expected, the obtained Husimi distribution is positive everywhere, hence it can be interpreted as a probability distribution of gluons in the Color Glass Condensate.

All the three distributions exhibit a peak in the $k$-direction and the peak location $k=Q_s(Y,b)$ or $k=Q_s(Y,\Delta)$ is identified with the saturation momentum.  It makes perfect sense that the phase space distributions in the saturated regime are characterized by the saturation momentum. As suggested recently \cite{Hatta:2016dxp} (see also \cite{Altinoluk:2015dpi}), the $k$-dependence of these distributions can be probed in diffractive dijet production in DIS where $k$ is correlated with the relative dijet momentum ${\bm P}_T = \frac{1}{2}(\bk_2 -\bk_1)$. Clearly one has to look at the region $P_T\sim Q_s$ in order to maximize the signal.

We have also extracted the elliptic part  which is also measurable in DIS  \cite{Hatta:2016dxp}. The peak moves at a slower speed  than in the angular independent part. This is because there is no geometric scaling in the elliptic part.   We also observed that the angular dependence is at most a few percent effect. Hopefully, the future Electron-Ion Collider (EIC) experiment \cite{Accardi:2012qut} is capable of detecting such a small effect.    


For phenomenological purposes, it is necessary to take into account higher order corrections to the BK equation with impact parameter dependence.  This was partly done in \cite{Berger:2010sh}, but more recently  generalizations of the BK equation which include the double-logarithmic resummation have been derived \cite{Beuf:2014uia,Iancu:2015vea}.  One cannot assume the SO(3) invariance anymore once these corrections have been included, although we suspect some remnant of the symmetry could survive.  Another direction is to include  the finite-$N_c$ corrections  by solving the Balitsky-JIMWLK equation \cite{Balitsky:1995ub,JalilianMarian:1997gr,Iancu:2000hn} and its collinearly improved version which resums double-logarithmic corrections \cite{Hatta:2016ujq}.
However,  solving the JIMWLK equation including the $\bb$-dependence appears to be a challenging task.


\section*{Acknowledgments}
We thank Edmond Iancu and Anna Stasto  for discussions and comments. 
The work of T.~U. is supported by the ERC Advanced Grant
  no. 320651, ``HEPGAME''. Numerical computations have been partly carried out at the Yukawa Institute Computer Facility.

\appendix
\section{Proof of $ d^2 \le 1$}\label{App_proofd2}
In this Appendix we prove that $d^2(\x,\y) \le 1$, or equivalently, 
\be
d^2(\x,\y) \le 1 &\iff & R^2(\x-\y)^2 \le (R^2+\x^2)(R^2+\y^2)  \nonumber \\
&\iff & 0 \le R^4+\x^2\y^2+2R^2(\x\cdot \y)\,.
\ee
Using the Cauchy-Schwartz inequality 
\be
(\x\cdot \y)^2 \le (\x)^2 (\y)^2 \,,
\ee
we find
\be
R^4+\x^2\y^2+2R^2(\x\cdot \y) &\ge & R^4+(\x\cdot \y)^2 + 2R^2(\x\cdot \y) \nonumber \\
&= & (R^2+(\x\cdot \y))^2\ge  0\,. \qquad \qquad ({\rm Q.E.D.})
\ee

\section{Alternative approach to solve the BK equation
}\label{another_parametrization}

Instead of setting $\y=0$ as in the main text, here let us set  $\y = -\x$ so that
\be
\label{dxx}
d^2(\x,-\x) =  \frac{4R^2\x^2}{(R^2+\x^2)^2} \,.
\ee
This function maps a finite interval $0\le |\x| \le R$ into $0 \le d^2 \le 1 $ monotonically, so it suffices to determine  $S_Y(\x,-\x)$ in the range $0\le |\x| \le R$. Let us therefore split the right hand side of the BK equation as
\be
\partial_Y S_Y(\x,-\x)=
\bar{\alpha}_s\left(\int_{|\z|<R} +\int_{|\z|>R}\right)\frac{d^2\z}{2\pi}\frac{4x^2}
{(\x-\z)^2(\z+\x)^2}\left(S_Y(\x,\z)S_Y(\z,-\x)-S_Y(\x,-\x)\right)\,. \nonumber \\
\label{ge}
\ee
Consider the region $|\z|\le R$ first.  For a pair of vectors $\x,\z$, we can find an associated vector $\x_I(\x,\z)$ such that 
\be
d^2(\x,\z) = \frac{R^2(\x- \z)^2}{(R^2+x^2)(R^2+z^2)} = \frac{4R^2x_I^2}{(R^2+x_I^2)^2} = d^2(\x_I,-\x_I)\,.
\ee
This can be solved as
\be
\x^2_I = R^2\left\{ -1+\frac{2}{d^2(\x,\z)} \pm \frac{2}{d^2(\x,\z)}\sqrt{1-d^2(\x,\z)} \right\} \,,
\ee
 where the minus sign should be taken to ensure that $|\x_I|\le R$. 
Next, the region $|\z|\ge R$ can be mapped to the region $|\z'|\le R$ using conformal symmetry. In the complex notation $\omega= x_1+ix_2$, the equation is invariant under $\omega\to - R^2/\omega$.  By choosing $\x=(x,0)$, we can rewrite the $|\z|\ge R$ part of (\ref{ge})  as
\be
\bar{\alpha}_s \int_{|\z'|<R}\frac{d^2\z'}{2\pi}\frac{4x'^2}
{(\x'-\z')^2(\z'+\x')^2}\left(S_Y(\x',\z')S_Y(\z',-\x')-S_Y(\x',-\x')\right)\,, 
\ee
where $\x' = (-\frac{R^2}{x},0)$. Writing $S_Y(\x,-\x)\equiv h_Y(x)$, the equation takes the form 
\be
\partial_Y h_Y(x)
&=& \bar{\alpha}_s \int_{|\z|<R} \frac{d^2\z}{2\pi} \biggl\{ \frac{4x^2}{(\x-\z)^2(\z+\x)^2}\left(h_Y(x_I(\x,\z))h_Y(x_I(\z,-\x))-h_Y(x)\right)  \nonumber \\ 
&& \quad + \frac{4\x'^2}
{(\x'-\z')^2(\z'+\x')^2}\left(h_Y(x_I(\x',\z'))h_Y(x_I(\z',-\x'))-h_Y(x)\right)   \biggr\}\,,
\label{ha}
\ee
where in the last term we used
\be
x_I(\x',-\x') = \frac{R^4}{x'^2} = x^2\,.
\ee
(\ref{ha}) is slightly  more complicated than (\ref{gg}), but it has the advantage that the function $h_Y(x)$ is defined in the finite interval $0\le x\le R$. The S-matrix is then given by
\be
S_Y(\x,\y)=h_Y(x_I(\x,\y))\,.
\ee
We have checked that the solution obtained in this way is consistent with the one obtained from 
(\ref{gg}).


\begin{thebibliography}{99}\label{sec:TeXbooks}%



\bibitem{Ji:2003ak} 
  X.~d.~Ji,
  Phys.\ Rev.\ Lett.\  {\bf 91}, 062001 (2003)
  doi:10.1103/PhysRevLett.91.062001
  [hep-ph/0304037].

\bibitem{Belitsky:2003nz} 
  A.~V.~Belitsky, X.~d.~Ji and F.~Yuan,
  Phys.\ Rev.\ D {\bf 69}, 074014 (2004)
  doi:10.1103/PhysRevD.69.074014
  [hep-ph/0307383].

\bibitem{Lorce:2011kd} 
  C.~Lorc\'e and B.~Pasquini,
  Phys.\ Rev.\ D {\bf 84}, 014015 (2011)
  doi:10.1103/PhysRevD.84.014015
  [arXiv:1106.0139 [hep-ph]].




\bibitem{Lorce:2011ni} 
  C.~Lorc\'e, B.~Pasquini, X.~Xiong and F.~Yuan,
  Phys.\ Rev.\ D {\bf 85}, 114006 (2012)
  doi:10.1103/PhysRevD.85.114006
  [arXiv:1111.4827 [hep-ph]].

\bibitem{Mukherjee:2014nya} 
  A.~Mukherjee, S.~Nair and V.~K.~Ojha,
  Phys.\ Rev.\ D {\bf 90}, no. 1, 014024 (2014)
  doi:10.1103/PhysRevD.90.014024
  [arXiv:1403.6233 [hep-ph]].




\bibitem{Mukherjee:2015aja} 
  A.~Mukherjee, S.~Nair and V.~K.~Ojha,
  Phys.\ Rev.\ D {\bf 91}, no. 5, 054018 (2015)
  doi:10.1103/PhysRevD.91.054018
  [arXiv:1501.03728 [hep-ph]].


\bibitem{Liu:2015eqa} 
  T.~Liu and B.~Q.~Ma,
  Phys.\ Rev.\ D {\bf 91}, 034019 (2015)
  doi:10.1103/PhysRevD.91.034019
  [arXiv:1501.07690 [hep-ph]].

\bibitem{Chakrabarti:2016yuw} 
  D.~Chakrabarti, T.~Maji, C.~Mondal and A.~Mukherjee,
  Eur.\ Phys.\ J.\ C {\bf 76}, no. 7, 409 (2016)
  doi:10.1140/epjc/s10052-016-4258-7
  [arXiv:1601.03217 [hep-ph]].






\bibitem{Meissner:2009ww} 
  S.~Meissner, A.~Metz and M.~Schlegel,
  JHEP {\bf 0908}, 056 (2009)
  doi:10.1088/1126-6708/2009/08/056
  [arXiv:0906.5323 [hep-ph]].

\bibitem{Lorce:2013pza} 
  C.~Lorc\'e and B.~Pasquini,
  JHEP {\bf 1309}, 138 (2013)
  doi:10.1007/JHEP09(2013)138
  [arXiv:1307.4497 [hep-ph]].



\bibitem{Echevarria:2016mrc} 
  M.~G.~Echevarria, A.~Idilbi, K.~Kanazawa, C.~Lorc\'e, A.~Metz, B.~Pasquini and M.~Schlegel,
  Phys.\ Lett.\ B {\bf 759}, 336 (2016)
 doi:10.1016/j.physletb.2016.05.086
  [arXiv:1602.06953 [hep-ph]].




\bibitem{Hagiwara:2014iya} 
  Y.~Hagiwara and Y.~Hatta,
  Nucl.\ Phys.\ A {\bf 940}, 158 (2015)
  doi:10.1016/j.nuclphysa.2015.04.005
  [arXiv:1412.4591 [hep-ph]].

\bibitem{Hatta:2011ku} 
  Y.~Hatta,
  Phys.\ Lett.\ B {\bf 708}, 186 (2012)
  doi:10.1016/j.physletb.2012.01.024
  [arXiv:1111.3547 [hep-ph]].




\bibitem{Courtoy:2013oaa} 
  A.~Courtoy, G.~R.~Goldstein, J.~O.~Gonzalez Hernandez, S.~Liuti and A.~Rajan,
  Phys.\ Lett.\ B {\bf 731}, 141 (2014)
  doi:10.1016/j.physletb.2014.02.017
  [arXiv:1310.5157 [hep-ph]].

\bibitem{Kanazawa:2014nha} 
  K.~Kanazawa, C.~Lorc\'e, A.~Metz, B.~Pasquini and M.~Schlegel,
  Phys.\ Rev.\ D {\bf 90}, no. 1, 014028 (2014)
  doi:10.1103/PhysRevD.90.014028
  [arXiv:1403.5226 [hep-ph]].

\bibitem{Rajan:2016tlg} 
  A.~Rajan, A.~Courtoy, M.~Engelhardt and S.~Liuti,
  arXiv:1601.06117 [hep-ph].


\bibitem{optics} D. T. Smithey, M. Beck, M. G. Raymer, and A. Faridani, Phys. Rev. Lett. 70, 1244  (1993). 


\bibitem{Hatta:2016dxp} 
  Y.~Hatta, B.~W.~Xiao and F.~Yuan,
  Phys.\ Rev.\ Lett.\  {\bf 116}, no. 20, 202301 (2016)
  doi:10.1103/PhysRevLett.116.202301
  [arXiv:1601.01585 [hep-ph]].


\bibitem{Altinoluk:2015dpi} 
  T.~Altinoluk, N.~Armesto, G.~Beuf and A.~H.~Rezaeian,
  Phys.\ Lett.\ B {\bf 758}, 373 (2016)
  doi:10.1016/j.physletb.2016.05.032
  [arXiv:1511.07452 [hep-ph]].

\bibitem{Balitsky:1995ub} 
  I.~Balitsky,
  Nucl.\ Phys.\ B {\bf 463}, 99 (1996)
  doi:10.1016/0550-3213(95)00638-9
  [hep-ph/9509348].

\bibitem{Kovchegov:1999yj} 
  Y.~V.~Kovchegov,
  Phys.\ Rev.\ D {\bf 60}, 034008 (1999)
  doi:10.1103/PhysRevD.60.034008
  [hep-ph/9901281].


\bibitem{GolecBiernat:2003ym} 
  K.~J.~Golec-Biernat and A.~M.~Stasto,
  Nucl.\ Phys.\ B {\bf 668}, 345 (2003)
  doi:10.1016/j.nuclphysb.2003.07.011
  [hep-ph/0306279].


\bibitem{Ikeda:2004zp} 
  T.~Ikeda and L.~McLerran,
  Nucl.\ Phys.\ A {\bf 756}, 385 (2005)
  doi:10.1016/j.nuclphysa.2005.03.119
  [hep-ph/0410345].


\bibitem{Marquet:2005zf} 
  C.~Marquet and G.~Soyez,
  Nucl.\ Phys.\ A {\bf 760}, 208 (2005)
  doi:10.1016/j.nuclphysa.2005.05.198
  [hep-ph/0504080].
  
\bibitem{Berger:2010sh} 
  J.~Berger and A.~Stasto,
  Phys.\ Rev.\ D {\bf 83}, 034015 (2011)
  doi:10.1103/PhysRevD.83.034015
  [arXiv:1010.0671 [hep-ph]].




\bibitem{Gubser:2011qva} 
  S.~S.~Gubser,
  Phys.\ Rev.\ D {\bf 84}, 085024 (2011)
  doi:10.1103/PhysRevD.84.085024
  [arXiv:1102.4040 [hep-th]].



\bibitem{Gelis:2010nm} 
  F.~Gelis, E.~Iancu, J.~Jalilian-Marian and R.~Venugopalan,
  Ann.\ Rev.\ Nucl.\ Part.\ Sci.\  {\bf 60}, 463 (2010)
  doi:10.1146/annurev.nucl.010909.083629
  [arXiv:1002.0333 [hep-ph]].

\bibitem{Hatta:2009nd} 
  Y.~Hatta and T.~Ueda,
  Phys.\ Rev.\ D {\bf 80}, 074018 (2009)
  doi:10.1103/PhysRevD.80.074018
  [arXiv:0909.0056 [hep-ph]].

\bibitem{Bondarenko:2015fca} 
  S.~Bondarenko and A.~Prygarin,
  JHEP {\bf 1506}, 090 (2015)
  doi:10.1007/JHEP06(2015)090
  [arXiv:1503.05437 [hep-ph]].



\bibitem{Hatta:2007fg} 
  Y.~Hatta and A.~H.~Mueller,
  Nucl.\ Phys.\ A {\bf 789}, 285 (2007)
  doi:10.1016/j.nuclphysa.2007.03.003
  [hep-ph/0702023 [HEP-PH]].

\bibitem{Gribov:1984tu} 
  L.~V.~Gribov, E.~M.~Levin and M.~G.~Ryskin,
  Phys.\ Rept.\  {\bf 100}, 1 (1983).
  doi:10.1016/0370-1573(83)90022-4


\bibitem{Iancu:2002tr} 
  E.~Iancu, K.~Itakura and L.~McLerran,
  Nucl.\ Phys.\ A {\bf 708}, 327 (2002)
  doi:10.1016/S0375-9474(02)01010-2
  [hep-ph/0203137].



\bibitem{Accardi:2012qut} 
  A.~Accardi {\it et al.},
  Eur.\ Phys.\ J.\ A {\bf 52}, no. 9, 268 (2016)
  doi:10.1140/epja/i2016-16268-9
  [arXiv:1212.1701 [nucl-ex]].


\bibitem{Beuf:2014uia} 
  G.~Beuf,
  Phys.\ Rev.\ D {\bf 89}, no. 7, 074039 (2014)
  doi:10.1103/PhysRevD.89.074039
  [arXiv:1401.0313 [hep-ph]].

\bibitem{Iancu:2015vea} 
  E.~Iancu, J.~D.~Madrigal, A.~H.~Mueller, G.~Soyez and D.~N.~Triantafyllopoulos,
  Phys.\ Lett.\ B {\bf 744}, 293 (2015)
  doi:10.1016/j.physletb.2015.03.068
  [arXiv:1502.05642 [hep-ph]].


\bibitem{JalilianMarian:1997gr} 
  J.~Jalilian-Marian, A.~Kovner, A.~Leonidov and H.~Weigert,
  Phys.\ Rev.\ D {\bf 59}, 014014 (1998)
  doi:10.1103/PhysRevD.59.014014
  [hep-ph/9706377].

\bibitem{Iancu:2000hn} 
  E.~Iancu, A.~Leonidov and L.~D.~McLerran,
  Nucl.\ Phys.\ A {\bf 692}, 583 (2001)
  doi:10.1016/S0375-9474(01)00642-X
  [hep-ph/0011241].



\bibitem{Hatta:2016ujq} 
  Y.~Hatta and E.~Iancu,
  JHEP {\bf 1608}, 083 (2016)
  doi:10.1007/JHEP08(2016)083
  [arXiv:1606.03269 [hep-ph]].




\end{thebibliography}
\end{document}